\documentclass[10pt,conference]{IEEEtran}
\IEEEoverridecommandlockouts
\usepackage{cite}
\usepackage{amsmath,amssymb,amsfonts}
\usepackage{algorithmic}
\usepackage{graphicx}
\usepackage{textcomp}
\usepackage{xcolor}

\def\BibTeX{{\rm B\kern-.05em{\sc i\kern-.025em b}\kern-.08em
    T\kern-.1667em\lower.7ex\hbox{E}\kern-.125emX}}
\begin{document}

\font\myfont=cmr12 at 18pt

\title{\myfont Automated Self-Admitted Technical Debt Tracking at Commit-Level: A Language-independent Approach}


\author{\IEEEauthorblockN{ Mohammad Sadegh Sheikhaei}
\IEEEauthorblockA{\textit{School of Computing, Queen's University} \\
Kingston, ON, Canada \\
sadegh.sheikhaei@queensu.ca}
\and
\IEEEauthorblockN{ Yuan Tian}
\IEEEauthorblockA{\textit{School of Computing, Queen's University} \\
Kingston, ON, Canada \\
y.tian@queensu.ca}
}

\maketitle

\begin{abstract}
Software and systems traceability is essential for downstream tasks such as data-driven software analysis and intelligent tool development. However, despite the increasing attention to mining and understanding technical debt in software systems, specific tools for supporting the track of technical debts are rarely available. In this work, we propose the first programming language-independent tracking tool for self-admitted technical debt (SATD) -- a sub-optimal solution that is explicitly annotated by developers in software systems. Our approach takes a git repository as input and returns a list of SATDs with their evolution actions (created, deleted, updated) at the commit-level. Our approach also returns a line number indicating the latest starting position of the corresponding SATD in the system. Our SATD tracking approach first identifies an initial set of raw SATDs (which only have created and deleted actions) by detecting and tracking SATDs in commits' hunks, leveraging a state-of-the-art language-independent SATD detection approach. Then it calculates a context-based matching score between pairs of deleted and created raw SATDs in the same commits to identify SATD update actions. The results of our preliminary study on Apache Tomcat and Apache Ant show that our tracking tool can achieve a F1 score of 92.8\% and 96.7\% respectively. 

\end{abstract}

\begin{IEEEkeywords}
SATD, technical debt, software tracing, technical debt tracking, SATD management
\end{IEEEkeywords}

\maketitle

\section{Introduction}\label{sec:introduction}

Self-admitted technical debt (SATD) refers to the sub-optimal design and implementation decisions that developers explicitly acknowledge~\cite{potdar2014}. A sample comment indicating the occurrence of a SATD is: ``\textit{TODO: - This method is too complex, let's break it up}''. Existing studies on SATDs mainly focus on understanding the prevalence of SATDs in specific software systems~\cite{potdar2014,yasmin2022first}, and how SATDs get removed/re-payed~\cite{maldonado2017empirical,liu2021AnES}. However, little is known about the updates of SATDs. Moreover, many SATD removal identified by comparing the content of SATD comments across versions are false alarms~\cite{zampetti2018was}.

To help developers and researchers properly link SATD instances across versions, a specific SATD tracking tool is essential. As far as we know, only one such tool exists, i.e., the SATDBailiff proposed by Alomar et al.~\cite{alomar2022satdbailiff}. SATDBailiff can track actions on a SATD instance from the history of a git repository, including created, deleted, and various types of updates, such as changing the name of the class/method/file path containing the SATD and changing the description of the SATD. It is reported to have 99\% precision in identifying SATD removal and 88\% for SATD update. SATDBailiff is designed for software written in Java, as it needs a Java parser to extract the class/method containing each SATD. This makes SATDBailiff inaccessible for software systems written in Python and many other programming languages. Moreover, while the precision of SATD removal and update action is high, the recall could be low. It should be noted that these performance values are action-level performance values. The SATD-level performance values would be lower, as any misidentified actions of a SATD would result in the SATD not being tracked correctly.

Unlike SATDBailiff, we aim to develop the first language-independent SATD tracking tool. Our tool leverages a novel context-based SATD matching algorithm that can map SATDs across versions without the help of a specific parser. Specifically, our approach contains three steps. In the first step, we extract the metadata of commits. In the second step, we extract the raw SATDs by detecting and tracking SATDs in commit hunks, leveraging a state-of-the-art language-independent SATD detection approach~\cite{guo2021far}. The raw SATDs only contain creation and deletion actions, and contain many false positives that appear as pairs of deletion/creation, but are actually update actions. Finally, in the third step, we fix the raw SATDs and their actions by identifying the false positive pairs of deletion/creation and treating them as the update actions for the merged raw SATDs. A preliminary evaluation on two well-known Apache projects, i.e., Tomcat and Ant, shows that our tracking approach can achieve an F1 score of 92.8\% and 96.7\%, respectively, which also outperforms SATDBailiff. 

\vspace{0.1cm}
\noindent \textbf{Main contributions:} Researchers can leverage our tool\footnote{Available from Figshare at DOI: 10.6084/m9.figshare.22229575} for fine-grained analyses of SATD lifespans in software systems written in different programming languages. Developers can utilize our tool to track and monitor SATDs along with software evolution.

\section{Background and Related Work}\label{sec:backgroud}

\subsection{SATD Identification and Tracking}

SATD identification is one of the primary steps in studying self-admitted technical debt. Several approaches have been proposed to detect SATD, from pattern-based methods~\cite{potdar2014,guo2021far} to supervised learning models~\cite{huang2018identifying,liu2018satd}. Potdar and Shihab~\cite{potdar2014} proposed 62 patterns (i.e., keywords and phrases) to detect SATD. Examples of the keywords/phrases are \textit{hack, fixme, is problematic, this isn’t very solid, probably a bug}.

Guo et al.~\cite{guo2021far} state that existing approaches proposed for SATD identification have neglected the fact that in many real projects, many comments have already been marked with various task annotation tags, predefined either by popular IDEs or by developers, to indicate SATD. They proposed a simple heuristic approach, named \textit{MAT}, that fuzzily matches task annotation tags which can achieve similar or even superior performance for SATD identification compared with existing approaches. We leverage their simple heuristic approach to identify SATD in this study because they are more efficient, language-independent, and have proved to have comparable performance with other complex SATD identification approaches.

The most closely related work is \textit{SATDBailiff}, a SATD mining and tracking tool proposed by Alomar et al.~\cite{alomar2022satdbailiff}. SATDBailiff is designed for Java projects and it is built upon an existing state-of-the-art SATD detection tool~\cite{huang2018identifying}. We have mentioned the limitations of SATDBailiff in Section~\ref{sec:introduction}. A preliminary comparison between our language-independent approach and SATDBailiff is presented in Section~\ref{sec:results}.

\subsection{vcsSHARK}

vcsSHARK~\cite{vcsSHARK} is a highly extensible repository parser developed by Trautsch et al.~\cite{Trautsch2017}, as a component of their SmartSHARK~\cite{smartshark} platform for supporting the replicability and validity of repository mining studies. vcsSHARK collects the complete commit history of a project including, e.g., all commits on all branches, tags, differences between file revisions, and changed files. The collected data is then stored in a MongoDB database. Our tool is build upon vcsSHARK, i.e., we utilize vscSHARK to collect structured commit information from a given git repository. 

\section{Approach}\label{sec:method}

Our approach contains the following three steps:
\begin{itemize}
    \item \textbf{Step 1:} Collect meta information of commits using vcsSHARK.
    \item \textbf{Step 2:} Identify and track raw SATDs from commit hunks. In this step, we extract all SATD creation and deletion actions, while some pairs of deletion/creation which occurred in the same file and commit are ``SATD update action" rather ``resolving a SATD and creating another SATD". We refer to these pairs as false positives.
    \item \textbf{Step 3:} Convert false positive SATD deletion/creation actions to SATD update actions.
\end{itemize}

Table~\ref{table-sample-SATDs-in-Tomcat} demonstrates an example output of our approach. Next, we describe the details of each step in the following subsections.

\begin{table*}[h]
\caption{Simplified sample tracked SATDs from Tomcat project}
\vspace*{-0.2cm}
\centering
\resizebox{18cm}{!}{%
\begin{tabular}{|c|c|c|c|c|c|c|c|c|}
\hline
\textbf{created} & \textbf{last appeared} & \textbf{created} & \textbf{last appeared} & \textbf{created} & \textbf{deleted} & \textbf{line content when created} & \textbf{updated} & \textbf{updated} \\
\textbf{in file} & \textbf{in file} & \textbf{in line} & \textbf{in line} & \textbf{in commit} & \textbf{in commit} & \textbf{line contents after update} & \textbf{in lines} & \textbf{in commits} \\
\hline 
IOChannel.java & IOChannel.java & 39 & 74 & e2f543b515 & c77f96a0bd & // TODO: update, etc & [(39,74)] & [fb3b1a48c6] \\
 &  &  &  &  &  & // TODO: update and use it (placeholder) &  &  \\
\hline
TesterWsCloseClient.java & TesterWsClient.java & 43 & 43 & 7b748998da &  & // TODO: Hoping this causes ... &  &  \\
\hline
\end{tabular}
\label{table-sample-SATDs-in-Tomcat}
}
\end{table*}

\subsection{Commits Extraction}

Our SATD tracking approach is based on the information extracted by running the vcsSHARK tool on a git repository. To identify and track SATDs, we use the following information returned by vcsSHARK:
\begin{itemize} 
    \item \textbf{project:} contains the main information of each repository. 
    
    \item \textbf{branch:} contains branch-related information, such as the id, project, and an is\_origin\_head field that determines if it is the master branch.

    \item \textbf{commit:} contains commit-related information, such as the id, project, branches, date, commit\_sha, and parents. 

    \item \textbf{file:} contains information related to each file, such as the id, path, and project. If a file path changes (e.g. a file is renamed or moved to another directory), a new document (row) is created in this table.

    \item \textbf{file\_action:} contains information about the different actions done on each file in each commit. The features include id, commit\_id, file\_id, mode (A=added, D=deleted, M=modified, C=copy-edit, R=rename-edit, U=unmerged), and old\_file\_id (if file renamed or copied). 

    \item \textbf{hunk:} contains the new added lines and deleted lines for a specific file\_action. The hunk features include id, file\_action\_id, content (that shows added and deleted lines), new\_lines (number of new lines), new\_start (the line number of the file where the new content is added), old\_lines (number of deleted lines), and old\_start (the line number of the file where the deleted lines start).
 
\end{itemize}


\subsection{Identify and track raw SATDs from commit hunks}

The following steps show how we use different information stored in the MongoDB to identify and track raw SATDs.

\vspace{.1cm}
\noindent \textbf{Extract sequence of commits in the master branch:} We first extract the ordered sequence of commits in the master branch of the selected project. To achieve this goal, according to the commit dates, we find the last master branch commit of the project. Then, we find the sequence of commits in the master branch by moving backward from the last commit to the first commit. The ``parents'' feature is used to find the previous commit of each commit. If a commit has two or more parents, we take the one in the master branch.

\vspace{.1cm}
\noindent \textbf{Extract ordered list of file actions and hunks for each file:} Leveraging the extracted commit sequence, we extract an ordered sequence of file actions for each file that ever exist in the project, including the ones deleted before and in the last commit. If a file is renamed, all file actions on the old file\_id are moved to the new file\_id. Having the ordered list of file actions for each file, we extract the ordered list of hunks for each file.

\vspace{.2cm}
\noindent \textbf{Identify and track raw SATDs:} We scan the ordered list of hunks for each file to identify and track SATDs. We use the MAT approach~\cite{guo2021far} to find the lines that include a SATD. The MAT approach employs regular expressions to detect TODO/FIXME/XXX/HACK in code comments. For each found SATD in added lines in hunks (that starts with a ‘+’ sign), we store its file\_id, created\_in\_commit, created\_in\_hunk, created\_in\_line, and current\_line.

To track the identified SATDs, while we pass over the ordered list of hunks, we update the current\_line for all still-alive SATDs found up to that commit. For example, assume that the current\_line of an existing SATD before commit x is 50, and there are two hunks in commit x on that file. One of the hunks deletes 3 lines starting from line 20 and adds 4 lines on line 22. The other hunk adds 10 lines on line 60. These hunks update the current\_line of that SATD to 51, because the first hunk adds one line (-3+4=1) before the SATD, however the second hunk took place after the SATD and has no effect on the current\_line of that SATD.

For each identified SATD in deleted lines of hunks (that starts with a ‘-’ sign), we search ``the line number of the deleted SATD in hunk" in current\_line of all still alive SATDs, and then store the deletion information in deleted\_in\_commit and deleted\_in\_hunk. We don’t need deleted\_in\_line feature, because the current\_line shows the last seen line number of a SATD (either it has been deleted or still exists). If a SATD is removed, we would not update its current\_line while passing over the rest hunks on the ordered list of hunks. We refer to the extracted SATDs in this step as \textit{raw SATDs} because there are some false positive SATD creation and deletion that needs to be resolved in the next step.


\subsection{Convert false positive SATD deletion/creation actions to SATD update actions}

\begin{figure}[htbp]
\centerline{\includegraphics[width=0.4\textwidth]{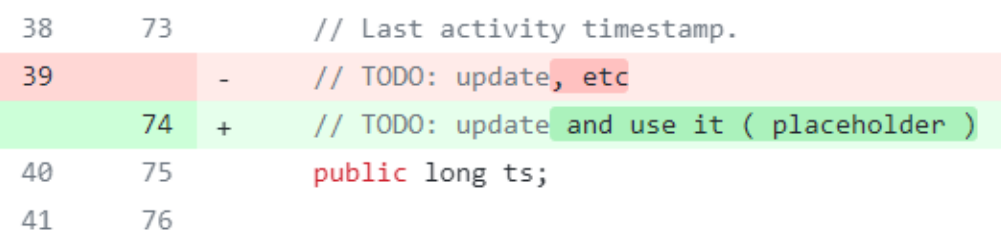}}
\vspace*{-3mm}
\caption{A false positive SATD deletion/creation: a TODO is deleted in line 39 and another TODO is created in line 74, while they both refer to the same SATD and it is just an update on TODO description to make it clear}
\label{fig-false-positive-example-2}
\end{figure}

A false positive SATD deletion/creation appears when a developer touches a SATD line without resolving that SATD. Figure~\ref{fig-false-positive-example-2} shows an example that the developer updates the description of a TODO without resolving it. As a result, in the extracted raw SATDs in the step 1, we observe a false SATD deletion and a false SATD creation action in that commit on that file, while both SATDs refer to the same SATD. In fact, no new SATD is created, but the description of that SATD is updated. On the other hand, it is possible that some SATDs are resolved by the developer, and some other new SATDs are created at the same commit on the same file. Therefore, for a specific commit and a specific file, we need to detect which SATD deletion/creation actions are true actions, and which pairs of deletion-creation are false positive.

To detect false positives in the raw SATDs, we designed a weighted multi-factor greedy algorithm that its weights are optimized by a grid search on manually labeled data. To describe this method, we define two phrases: ``\textit{following SATD candidate}" and ``\textit{following SATD}". A $rawSATD_j$ is a ``following SATD candidate" for $rawSATD_i$ if and only if the deletion of $rawSATD_i$ and the creation of $rawSATD_j$ occurred at the same commit and on the same file. A $rawSATD_j$ is the ``following SATD" of $rawSATD_i$ if and only if $rawSATD_j$ is a ``following SATD candidate" for $rawSATD_i$, and $rawSATD_j$ is an update on $rawSATD_i$, meaning both of them refer to the same SATD according to the content (SATD description) and context (surrounding code). 

The weighted multi-factor greedy algorithm gets the list of deleted SATDs and the list of created SATDs (all occurred in the same file and commit), and returns the following SATD (selected from created SATDs), if it exists, for each deleted SATD. We provide four features for each input SATD to the greedy algorithm: the SATD description, the previous line, the next line, and the hunk\_id of the SATD. We also provide the weight for each feature and a threshold that determines the minimum matching score. The weights and the threshold are obtained from the optimization process that is described later. The greedy algorithm applies the formulas in Table~\ref{table-matching-score-formula} to obtain the matching score between each pair of $deletedSATD_i$ and $createdSATD_j$.

\begin{table*}[h]
\caption{Calculating the matching score between $deletedSATD_i$ and $createdSATD_j$}
\vspace{-0.2cm}
\centering
\begin{tabular}{c}
\hline
$ matching\_score_{i,j}[feature] = jaccard\_similarity(deletedSATD_i[feature],\  createdSATD_j[feature]) $ \\
$ for \ feature \in \{description,\ previous\_line,\ next\_line\} $ \\
\\
$ matching\_score_{i,j}[hunk] = 1 \ if \ deletedSATD_i[hunk\_id] == createdSATD_j[hunk\_id], \ otherwise \ 0  $ \\
\\
$ matching\_score_{i,j} = \Sigma \ weight[feature] * matching\_score_{i,j}[feature] $ \\
$ for \ feature \in \{description,\ previous\_line,\ next\_line,\ hunk\} $ \\
\hline
\end{tabular}
\label{table-matching-score-formula}
\end{table*}

\begin{figure*}[t]
\centerline{\includegraphics[width=0.8\textwidth]{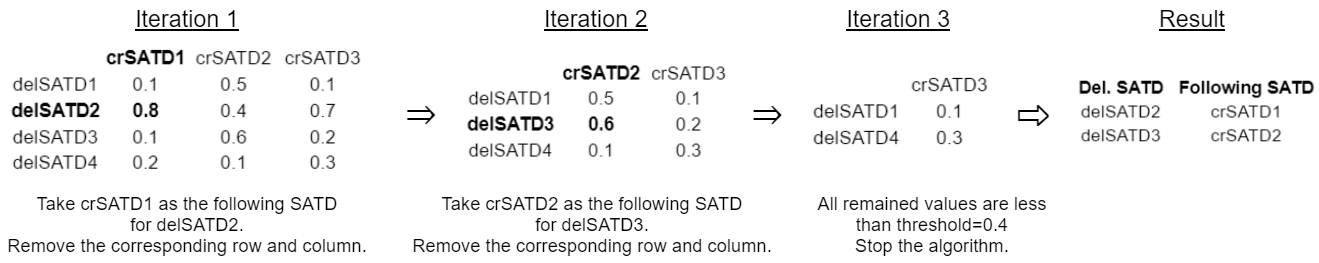}}
\caption{The iterations of the greedy algorithm on an example with four deleted and three created SATD to find the following SATDs for the deleted SATDs.}
\label{fig-greedy-algorithm-example}
\end{figure*}
The matching score is calculated using a combination of four factors (Table~\ref{table-matching-score-formula}): factors one to three are the similarity between the descriptions/previous\_line/next\_line of the deleted and created SATD, and the fourth factor is whether the hunk ID of the deleted and created SATDs match. The Jaccard similarity is used to calculate the similarity between the two descriptions, the two previous\_lines, or the two next\_lines, and a hunk ID match is given a score of 1. The final matching score is calculated by multiplying the factor similarity scores by their respective weights (description\_weight, prev\_line\_weight, next\_line\_weight, hunk\_weight) and then summing them together.

The greedy algorithm starts with creating the matching\_score matrix using the above mentioned formulas. It then repeatedly takes the highest value in the matrix and compares it against the given threshold to decide if the selected pair is a matching SATD. Before going to the next iteration and taking the next highest value, the algorithm removes the corresponding row and column of the highest value from the matrix to ensure each SATD is involved at most once in the list of matching SATDs. If the next highest value in the matrix is lower than the threshold, the algorithm stops the loop and returns the matched SATD pairs. Figure~\ref{fig-greedy-algorithm-example} shows the iterations of the greedy algorithm on an example with four deleted SATDs and three created SATDs all occurred in the same file and the same commit. At the first iteration, the algorithm takes crSATD1 as the following SATD for delSATD2, and removes their corresponding row and column from the matching score matrix. The iterations continues until iteration 3 that all matching scores are less than the threshold=0.4, and the algorithm returns the matched SATDs.

To prepare manually labeled data for our weight optimization process, we applied steps 1 and 2 on two randomly selected Apache projects, Commons Math\footnote{https://github.com/apache/commons-math} and ActiveMQ\footnote{https://github.com/apache/activemq}, to extract the raw SATDs, and then wrote a simple code to extract the ``following SATD candidates" for each raw SATD. Among 629 raw SATDs extracted from Commons Math project, 93 have at least one candidate. Similarly, out of 623 raw SATDs extracted from ActiveMQ project, 162 have at least one candidate. We then manually label each raw SATD with the true match of its following SATD candidates, if any, as its following SATD. To ensure the accuracy of the manual labeling, we consider the content and context of SATDs. Content matching refers to the comparison of the descriptions of the two SATDs, while context matching refers to the comparison of the code lines surrounding the two SATDs. For example, in Figure~\ref{fig-false-positive-example-2}, the content of the SATD has changed from ``TODO: update, etc" to ``TODO: update, and use it (placeholder)", which alone might not be enough to conclude that they refer to the same SATD. However, when we also consider the context, we see that the surrounding code lines are the same, which allows us to conclude that the created SATD is the following SATD of the deleted SATD.

To find the best weights and threshold for the greedy algorithm, we conducted a grid search by applying the greedy algorithm with all possible discrete weights and thresholds on the two projects that we manually generated their golden labels. For each four weights, we considered values of 0, 0.1, 0.2, …, 1.0, with a precondition to ensure the sum of weights equaled 1.0, and for the threshold we considered values of 0, 0.1, 0.2, …, 1.0. The optimum result of our grid search was obtained by description\_weight=0.6, prev\_line\_weight=0.2, next\_line\_weight=0, hunk\_weight=0.2, and the threshold=0.4. Using the optimum weights and threshold, the greedy algorithm achieved 97.5\% accuracy over both projects.

To create the final list of SATDs, we merge each raw SATD with its actual following SATD, if one exists. The merged SATD inherits its creation features from the deleted SATD, and inherits its deletion features from the following SATD. The features of the false positive deletion/creation pair (that all occurred at the same commit) would be considered as the update features for the merged SATD. If a raw SATD does not have a following SATD, it will be included in the final list as is. If a raw SATD is followed by another raw SATD, which is in turn followed by a third raw SATD, and so on, all of them will be merged into a single SATD (with several updates) in the final list. This process ensures that the final list of SATDs is an accurate representation of all distinct SATDs that were present in the target repository at different points in time.

\section{Preliminary Evaluation}\label{sec:results}
\subsection{Target Projects}
Using the vcsSHARK tool, we collect the data of two well known Apache projects, i.e., Tomcat\footnote{https://github.com/apache/tomcat} (collected in Sep 2022) and Ant\footnote{https://github.com/apache/ant} (collected in Apr 2022) from GitHub. Then we apply our algorithm to extract and track SATDs. Table~\ref{table-number-of-tracked-satd} shows the number of extracted raw SATDs (step 2), the final number of SATDs after removing false positives and merging raw SATDs (step 3), and the total number of false positives that indicates the number of SATD updates.

\vspace{-0.1cm}
\begin{table}[h]
\caption{Number of tracked SATDs for the selected two projects}\vspace{-0.2cm}
\centering
\begin{tabular}{|l|c|c|}
\hline
\textbf{} & \textbf{Tomcat} & \textbf{Ant} \\
\hline
Number of commits & 70,666 & 17,162 \\
Number of commits in the master branch & 24,069 & 14,366 \\
Number of raw SATDs (in step 2) & 3,370 & 1,171  \\
Number of final SATDs (in step 3) & 2,587 & 806  \\
Number of false positives (SATD updates) & 783 & 365  \\
\hline
\end{tabular}
\label{table-number-of-tracked-satd}
\end{table}

\subsection{Baseline and Evaluation Method}
Our baseline approach is SATDBailiff, but we made two modifications to make it comparable to our approach. Firstly, we replaced its SATD identification method with the MAT approach. Secondly, we altered the algorithm to only extract SATD actions from the commits in the master branch. We have named this modified version as SATDBailiff2.

We use a SATD-level evaluation method. The following definitions are used to calculate the standard classification evaluation metrics, including Precision, Recall, and F1-score: 

\begin{itemize}
    \item True Positive (TP): If the creation, deletion, and all update actions of an identified SATD is correct.
    \item False Positive (FP): If the creation, deletion, or any of update actions of an identified SATD is wrong.
    \item False Negative (FN): If any actions of a correctly identified SATD by the opponent approach is not identified correctly by this approach. 
\end{itemize}

The creation of an identified SATD is correct if and only if its created\_in\_line, created\_in\_file, and created\_in\_commit is correct, and there is no similar SATD in that specific file and commit that is deleted (otherwise it is an update action). Similarly, the deletion of an identified SATD is correct if and only if its deleted\_in\_line, deleted\_in\_file, and deleted\_in\_commit are correct, and there is no similar SATD in that specific file and commit that is created. An update action is considered correct if SATD A is deleted and SATD B is created in the same commit and file, and SATD B follows SATD A.

\subsection{Results}

Table~\ref{table-performance} compares the performance of our approach with SATDBailiff2 on two projects. As a preliminary evaluation, we selected SATDs with a creation line equals to 36 or less for the Tomcat project, and 56 or less for the Ant project.\footnote{These two specific location-related thresholds are set to ensure we have around 100 SATDs labeled for our approach and the baseline.} Among 104 extracted SATDs from Tomcat project by our approach, 100 are correct and 4 are incorrect due to miss-detecting the following SATDs. Among 94 extracted SATDs by SATDBailiff2, 87 are correct and 7 are incorrect. Among 87 correctly identified SATDs by SATDBailiff2, 9 SATDs missed by our approach. Among 100 correctly identified SATDs by our approach, 10 SATDs missed by SATDBailiff2. 

\vspace{-0.2cm}
\begin{table}[h]
\caption{The performance of our approach vs SATDBailiff2}\vspace{-0.2cm}
\centering
\begin{tabular}{|l|c c|c c|}
\hline
\textbf{} & \multicolumn{2}{c|}{Tomcat}  & \multicolumn{2}{c|}{Ant}  \\
\hline
 Approach & Our & SATDBailiff2 & Our & SATDBailiff2   \\
Precision & 96.2\% & 92.6\% & 97.7\% & 87.5\%   \\
Recall & 89.7\% & 90.0\% & 95.7\% & 78.8\%   \\
F1-score & 92.8\% & 91.3\% & 96.7\% & 82.9\%   \\
\hline
\end{tabular}
\label{table-performance}
\end{table}

\section{Conclusion and Future Work}\label{sec:conclusion}
In this paper, we proposed the first language-independent approach for tracking self-admitted technical debt (SATD) through commit hunks. To address the issue of SATD update actions being presented as deletion/creation action pairs in commit hunks, we proposed a multi-factor greedy algorithm to detect these pairs as false positives. We then merged the corresponding SATDs involved in the paired false positives and treated the false positive actions as their update actions. The evaluation results demonstrate the effectiveness of our proposed approach in accurately tracking SATDs across all commits in a repository. However, we also want to point out that the current evaluation is still preliminary, i.e., our reported results may not hold on to other OSS projects. Further qualitative and quantitative evaluations should be performed to mitigate the threats introduced by the selection of SATDs for evaluation and the parameter tuning process. 


\section*{Acknowledgement}
We acknowledge the support of the Natural Sciences and Engineering Research Council of Canada (NSERC), [funding reference number: RGPIN-2019-05071].

\bibliographystyle{IEEEtran}
\bibliography{main}

\end{document}